\documentclass[twocolumn,showpacs,preprintnumbers,amsmath,amssymb]{revtex4-1}

\usepackage{graphicx}
\usepackage{dcolumn}
\usepackage{bm}
\usepackage{color}

\begin{document}


\title{Spontaneous emergence of rogue waves in partially coherent waves:  a quantitative experimental comparison between hydrodynamics and optics}

\author{ R. El Koussaifi$^{1,2}$, A. Tikan$^{1,2}$, A. Toffoli $^3$, S. Randoux$^{1,2}$, P. Suret$^{1,2}$ \& M. Onorato$^{4,5}$}

\affiliation{$^1$  Laboratoire de Physique des Lasers, Atomes et Molecules, Universit\'e de Lille, UMR-CNRS 8523, 59655 Villeneuve d'Ascq, France}
\affiliation{$^2$ Centre d'Etudes et de Recherches Lasers et Applications (CERLA),  59655 Villeneuve d'Ascq, France}
\affiliation{$^3$  Department of Infrastructure Engineering, The University of Melbourne, Parkville, VIC 3010, Australia}
\affiliation{$^4$ Dipartimento di Fisica, Universit\`a degli Studi di Torino, 10125 Torino, Italy}
\affiliation{$^5$  Istituto Nazionale di Fisica Nucleare, INFN, Sezione di Torino, 10125 Torino, Italy}

\date{\today}

\begin{abstract}
  Rogue waves are extreme and rare fluctuations of the wave field that have been discussed in many physical systems. Their presence substantially influences the statistical properties of an incoherent wave field. 
 Their understanding is fundamental for the design of ships and offshore platforms. 
  Except for very particular meteorological conditions, waves in the ocean are characterised by the so-called JONSWAP (Joint North Sea Wave Project) spectrum. Here we compare two unique experimental results: the first one has been performed in a 270-meter wave tank and the other in optical fibers. In both cases, waves characterised by a JONSWAP spectrum and random Fourier phases have been launched at the input of the experimental device. The quantitative comparison, based on an appropriate scaling of the two experiments, shows a very good agreement between the statistics in hydrodynamics and optics. Spontaneous emergence of heavy tails in the probability density function of the wave amplitude is observed in both systems. The results demonstrate  the universal features of rogue waves and provide a fundamental and explicit bridge between two important fields of research. Numerical simulations are also compared with experimental results.
\end{abstract}

\maketitle

\section{Introduction}

Ocean waves are primarily forced by the wind. After being generated,
waves are subjected to nonlinear interactions, which transfer energy
to different modes and thus model the ocean wave spectrum. In the late
sixties an international team of scientists carried out a field
campaign with the aim of finding the shape of the wave spectrum
\cite{komen94}. They measured the surface elevation in thirteen
stations along 160 km in the North Sea. Frequency Fourier spectra were
then computed and fitted by an empirical formula, which is known as
the JONSWAP spectrum \cite{janssenbook04}. Except for particular
meteorological or bathymetric conditions, nowadays it is well accepted
by the oceanographic community that, to some extents, wind waves are
described by such spectrum.

If the phases of the Fourier components of the JONSWAP spectrum are randomly distributed, the surface elevation is characterized by a Gaussian distribution, according to the Central Limit Theorem. This implies that the envelope is Rayleigh distributed and the square of the envelope (i.e. the power) obeys the Exponential distribution (see section \ref{sec:statistics}). From a purely statistical point of view, extreme events may always take place albeit seldomly. Just to give an example, according to the Rayleigh distribution, the probability of measuring a wave that is larger that 4 times the standard deviation of the surface elevation (a rogue wave) is 3.4$\times10^{-4}$. This implies that in a storm characterized by waves that have a mean frequency of 
0.1~Hz, it would take in principle 8.3 hours to measure such an extreme event (provided the meteo-ocean conditions remain stationary).

Due to nonlinearity, correlations of the phases can develop. Therefore, the statistical properties of the surface elevation may change and  rogue waves may appear more often than predicted by the linear theory. 
The origin of such waves is very much debated and different explanation may be found in the literature 
\cite{kharif2009rogue,onorato2013rogue,onorato2016onorigin,fedele2016real,akhmediev2009waves,dudley2014instabilities,onorato2016rogue}.
In the limit of weakly nonlinear one dimensional waves, simplified
forms of the primitive equation of motion may offer some insights on
the problem. After the pioneering work in \cite{TRUL97,henderson99},
it has become a common practice to investigate rogue waves using the
Nonlinear Schr\"odinger equation (NLSE). The motivation relies on the
fact that such equation has particular solutions   \cite{akhmediev1987exact,kuznetsov1977solitons,peregrine1983water, akhmediev2009waves, Akhmediev:13, dudley2014instabilities}, known as breathers, that resemble qualitatively the profile of rogue waves measured in the open ocean. 
Inspired by the work in oceanography, the concept of rogue wave  has
been developed in various optical systems \cite{ Solli:08,Erkintalo:09, Kibler:09, Mussot:09,Montina:09,Pierangeli:15,Bonatto:11,
Lecaplain:12,Kasparian:09,Hammani:08, Randoux:12,Walczak:15, Suret:16, Narhi:16}, starting from the
work in \cite{solli2007optical}. 
Experiments have also been performed in order to
generate exact solutions of the NLSE in optical fibers and water wave tanks
\cite{kibler2010peregrine,kibler2012observation,kibler2015superregular,kibler2016experiments,
Chabchoub:11,shemer2013peregrine, kibler2015superregular}. These
observations have been of fundamental significance for establishing
the bases of a potential bridge between the fields of optics and
hydrodynamics (see also \cite{chabchoub2015nonlinear,onorato2016hydrodynamic} for a detailed
comparison between the NLSE in optics and hydrodynamics). 
In all these optical fibers and water experiments, the initial
conditions are deterministic and specifically designed to match peculiar
exact solutions of  the NLSE.

A critical challenge in the field is the development of realistic
oceanic conditions in an optical fiber.  The accurate measurement of
the statistical properties of the wave amplitude in the fiber is also
a critical issue \cite{Jalali:10,Solli:12,Walczak:15,Suret:16, Narhi:16}. Typically ocean waves are far from being either monochromatic or a supercontinuum, but they are characterised by a finite width spectrum (partially coherent waves) in the weakly nonlinear regime. This is precisely the regime  investigated  in the present paper. 

Here we report an ad-hoc optical experimental set-up  that has been properly designed in order to propagate in an optical fiber waves characterised by a JONSWAP spectrum. We quantitatively compare the evolution of the statistical properties of waves in an optical fiber against the ones recorded in the long water wave tank at Marintek, Trondheim (Norway) \cite{ONO06,onoratoetal04}. 

In the optical experiment, the scaling of power and spectral width allows to investigate  several values of the normalized propagation length by using a given length of optical fiber. In particular this allowed us to study the statistics as a function of the propagation distance without the use of the so-called ``cut-back'' technique~\cite{Barviau:06,kibler2012observation, mussot2014fermi, Tikan:17}.

The paper is organized as follows: in Section \ref{sec:unifying} we present a unifying description of water and optical waves in the framework of NLSE; this is a very important step which allows us to design properly the experiment and scale the parameters in the optical experiment in order to match the hydrodynamical one. Indeed, we stress from the beginning that it is not our goal to establish quantitatively the validity of the NLSE in the respective fields or to model properly the complicated phenomena (wave breaking, friction etc.) that may take place in a wave tank. The role played here by the NLSE is to establish the spatial  and time scales over which comparison between optical and hydrodynamical experiments can be made.  In Section \ref{sec:statistics} we describe some basic features of the statistical properties of linear partially coherent waves and we define the observables that will be extracted from the measurements. For the sake of completeness, in Section \ref{sec:JONS} a quick formulation of the JONSWAP spectrum is given.
In Section \ref{sec:setup} both experimental set-ups will be described and results will be presented in Section
\ref{sec:results}. Conclusions will follow.

\section{A unifying description } \label{sec:unifying}
Wind waves have a typical spatial scale, $\lambda_0$, that ranges from a few centimeters to almost half kilometre. Those lengths corresponds approximately to frequencies of 5 Hz to 5$\times 10^{-2}$ Hz. Independently of the frequency of the carrier wave, $f_0$, the typical frequency spectral bandwidth, $\Delta f/f_0$, is of the order of 0.3. Due to the relatively small wave tank (280 meters), in our experiments we have considered waves characterized by a period of 1.5 seconds ($f_0\sim$ 0.667 Hz, $\lambda_0\sim$3.5 $m$). In optical fibers experiments based on telecommunications equipment, the wavelength of the carrier wave $\lambda_0\sim1.55$ $\mu$m that corresponds to $f_0\sim 2 \times 10^{14}$ Hz. In optical fiber experiments involving partially coherent waves described by the NLSE, the spectral bandwidths are typically  of the order of $0.5 \times10^{-3}$ \cite{Randoux:14, Walczak:15, Randoux:16,Suret:16}. 

 At first glance, one would be tempted to state that, due to the very different spatial and temporal scales, one would need a very short optical fiber to reproduce 
the wave dynamics in the water tank. As it will be clear soon, it turns out that the length of the 
fiber is of the same order as the length of the water wave tank. 

A simple and straightforward way for appreciating this fact can be 
deduced by analyzing the NLSE both in optics and in water waves.
While NLSE provides only an approximation to the dynamics of water waves, we have found that its use has been of practical relevance for the design of the optical experiment and for the comparison between the hydrodyanmical and optical data sets. Indeed, the NLSE equation offers a common background over which nonlinear dynamics in different fields can be described \cite{chabchoub2015nonlinear,onorato2016rogue}. 
 For the present discussion, it is important to write the NLSE in the following form:
 \begin{equation}
i\frac{\partial A}{\partial z}= \frac{1}{2}\beta_2\frac{\partial^2 A}{\partial t^2}-\chi^{(3)}  |A|^2A, \label{nls}
\end{equation}
where $z$ is the propagation variable, $\beta_2$ and $\chi^{(3)}$ are two known constant coefficients that account for the group dispersion and the strength of nonlinearity,
respectively. In water waves in infinite deep water $\chi^{(3)}=-k_0^3$
and $\beta_2=2/g$, with $g$ the acceleration of gravity and $k_0$ the
wave number of the carrier wave; in optics, in the anomalous
dispersion regime (the one investigated here), $\beta_2$ is negative
while $\chi^{(3)}$ is always positive in fibers. 
 It has to be mentioned that the surface elevation $\eta(x,t)$ is related at the leading order to the  slowly varying complex envelope $A$ as:
\begin{equation}
\eta(z,t)=\frac{1}{2}\left( A(z,t)e^{i (k_0z -\omega_0 t)}+c.c.\right); \label{etaA}
\end{equation}
a similar relation holds for the optical (electric) field. 

From (\ref{etaA}) the following relation can be derived :
\begin{equation}
\langle|A|^2\rangle=2\langle\eta^2\rangle=2 \sigma^2, \label{amplitudesurface}
\end{equation}
where $\langle...\rangle$ implies averages over time and $\sigma^2$ is the variance of the rapidly oscillating wave field.
While the surface elevation is directly measured in standard water wave experiments at fixed values of $z$, in optical experiments the quantity that is measured at the end of the fiber is the power which is proportional to the modulus square of the complex envelope $A$.

In order to design the experiment, it is useful to introduce from equation 
(\ref{nls}) a linear and a nonlinear propagation length  in hydrodynamics and in optics as follows:

\begin{equation}
z_{lin}=\frac{2}{\beta_2\Delta \omega^2}\;\;\ {\rm and}\;\;\;\;
z_{nlin}=\frac{1}{\chi^{(3)}\langle|A_0|^2\rangle},
\label{eq:zlznl}
\end{equation}
where $\Delta \omega=2\pi \Delta f$ is a typical spectral bandwidth  and $\langle|A_0|^2\rangle$ is the average value of 
the envelope square both calculated at $z=0$ that in optics
corresponds for instance to the average power $P_0$ injected in the fiber.

In oceanography the estimation of a characteristic amplitude  is
usually made by introducing the so called significant wave height,
$H_s$, that is defined as the average over 1/3 of the highest waves
in the measured time series  (a wave height is the distance between a
crest and the adjacent  trough).  Assuming Gaussian statistics for $\eta$,  $H_s\simeq 4\sigma$, with $\sigma=\sqrt{\langle \eta^2\rangle}$ being the standard deviation of the surface
elevation \cite{kharif2009rogue}. Using equation (\ref{amplitudesurface}), the following relation, connecting the power to the significant wave height, can be obtained directly:
\begin{equation}
  \langle|A_0|^2\rangle=P_{0}=H_s^2/8.
  \label{eq:P0ww}
\end{equation}
The degree of nonlinearity of the wave propagation is given by the parameter:
\begin{equation} 
\epsilon=\frac{z_{lin}}{z_{nlin}}=\frac{2 \chi^{(3)}\langle|A_0|^2\rangle}{\beta_2\Delta \omega^2}.
\label{eq:epsilon}
\end{equation}

In the context of ocean waves, $\sqrt{\epsilon}$ has been named as the Benjamin-Feir Index \cite{janssen03,onorato01} and it has been shown that when $\epsilon$ is large then the evolution will eventually lead to the formation of rogue waves with a probability larger than the expected from a Rayleigh distribution of the envelope.

 For a given value of $\epsilon$ the dynamics is unique, i.e., if we assume that waves propagate according to the NLSE, we expect to observe the same phenomenology in 
 the optical fiber and in the water tank. Varying the power $P_0= \langle|A_0|^2\rangle$ and accordingly the spectral width $\Delta \omega = 2 \pi \Delta f$ allows one to change  the nonlinear length while keeping constant $\epsilon$. This procedure can be used to obtain the desired nonlinear length at the end of the fiber, while keeping unchanged the dynamics. 
The last helpful ingredient for the comparison of the two experiments is the introduction of the time scale $\tau$ of the coherent structures such as solitons that are solutions of the NLSE. This is obtained by balancing the nonlinear and dispersive terms in equation (\ref{nls}):
\begin{equation}
\tau=\sqrt{\frac{|\beta_2|}{2 \langle|A_0|^2\rangle \chi^{(3)}}} 
\label{time}
\end{equation}
By considering this quantity as the time unit, we will be able to use dimensionless frequencies in the comparison of the spectra from the optical and hydrodynamical experiments.

\section{Statistical property of random waves: some basics properties} \label{sec:statistics}
If one assumes that the Fourier phases of the surface elevation or the electric filed  are uniformly distributed, then the 
probability density function (PDF) of the field is Gaussian. In optics the measurement of the electric field in time is not feasible and only the envelope can be measured (see  \cite{Suret:16,Narhi:16} for new developments). In order to make comparison between optics and hydrodynamics, it is then necessary to 
build the envelope from the surface elevation. The procedure is well known in the literature (see for example \cite{janssen2014random}) and consists in constructing a 
synthetic field, $\tilde\eta(t)$ (we consider it at fix $z$) that is orthogonal to the $\eta(t)$ and build the auxiliary complex variable
 $g(t)=\eta(t)+i \tilde\eta(t)$.  The variable $\tilde\eta(t)$ can be computed from the Hilbert Transform, which rotate the Fourier coefficients of $\eta(t)$ by $-\pi/2$  for positive frequencies and $\pi/2$ for negative frequencies. The modulus of $g(t)$ corresponds  to the modulus of  $A(t)$.
 Now assuming that $\eta(t)$ is the superposition of sinusoidal waves with random phases, then its probability 
 density function, $p(\eta)$, is Gaussian:
 \begin{equation}
 p(\eta)=\frac{1}{\sigma \sqrt{2 \pi}} \exp[{-\eta^2/(2\sigma^2)}],
 \end{equation}
 where $\eta(t)$ and $\tilde\eta(t)$ are two random variables.

 Assuming that they are independent variables with the same variance,
  then  the joint probability density function is given by:
  \begin{equation}
  p(\eta,\tilde\eta)= p(\eta)p(\tilde\eta)=\frac{1}{\sigma^2 2 \pi} \exp[{-(\eta^2+\tilde\eta^2)/(2\sigma^2)}].\label{eq:etatilde}
  \end{equation}
Considering the above probability density function in polar coordinates $(|A|, \theta)$ with $|A|=\sqrt{\eta^2+\tilde\eta^2}$ and $\theta$ the phase, and integrating over the values of  $\theta$  it is straightforward to show that
\begin{equation}
p(|A|)=\frac{|A|}{\sigma^2}\exp[-|A|^2/(2\sigma^2)]; \label{eq:rayleigh}
\end{equation}
this is the well known  Rayleigh distribution.  The fourth-order normalized  moment of such distribution is given by:
\begin{equation}
  \kappa=\frac{ \langle|A|^4\rangle}{\langle|A|^2\rangle^2}=\frac{\int_0^{\infty}p(|A|) |A|^4dA}{\left(\int_0^{\infty}p(|A|) |A|^2dA\right)^2}=2.
  \label{eq:kappa}
\end{equation}
Here $\kappa$ is one of the  observables which will be compared in optical and hydrodynamical experiments. Values larger that  2 imply that the tail of the distribution is fatter than the Rayleigh distribution, i.e. more rogue waves than predicted by linear theory should appear. 
Starting from (\ref{eq:rayleigh}), the probability density function
for the normalized intensity (or power)  $I=|A|^2/P_0 =|A|^2/(2\sigma^2)$ (see 
equation (\ref{amplitudesurface})) can be derived:
\begin{equation}
p(I)=\exp[-I]; \label{eq:exponential}
\end{equation}
The exponential distribution $p(I)$ will be our reference one, when comparing the optical and the hydrodynamical results.

In the typical experiments that will be discussed, at the beginning of the wave tank or optical fiber, we prescribe a spectral shape and random  phases. Therefore, very close to the inlet we expect to see $\kappa\simeq2$ and an exponential distribution for the intensity.  Afterwards, waves travel along the tank/fiber and, because of the nonlinearity, their spectrum changes and the statistical distribution of the wave intensity change as wel
\cite{onorato2016onorigin}, as we will discussed below.

 \section{The JONSWAP spectrum} \label{sec:JONS}
The Joint North Sea Wave Project took place in the late sixties in the North Sea 
\cite{hasselmann1973measurements}. Wave spectra were computed from 13 measurement stations 
over 160 kilometers. One of the aim of the experiment was to understand the nonlinear transfer in the energy balance equation. Observations  suggested that the spectral shape of the ocean waves depends on the stage of development of the sea state. In \cite{hasselmann1973measurements} the following parametrization of the frequency wave spectral density was proposed
\begin{equation}
S(f)= \frac{\alpha g^2}{(2 \pi)^4f^5}\exp\left[-\frac{5}{4}\left(\frac{f_0}{f}\right)^4\right]
\gamma^{\exp\left[-\frac{(f-f_0)^2}{2\tilde\sigma^2 f_0^2}\right]}, \label{eq:jonswap}
\end{equation}
with $g$ the gravity acceleration, $f_0$ the frequency corresponding to the peak of the spectrum, $\tilde\sigma=0.07$ if $f \le f_0$ and $\tilde\sigma=0.09$ if $f>f_0$;
$\alpha$ and $\gamma$ may assume different values depending on the sea state.
 As $\gamma$ increases, the spectrum becomes more narrow and 
the power also increases. Large values correspond to young seas, i.e. those for which the phase velocity of the waves is much smaller than the wind speed. The parameter $\alpha$ is related to the power: as it increases, the significant wave height
increases as well. As a measure of the width $\Delta f$ of the spectrum, we will use in the following the full width at half maximum.

Here we remark that the dynamics of nonviscous surface gravity 
 waves is fully scalable in the laboratory, in the sense that the dynamics of the 
ocean waves can be reproduced in a wave tank, provided the adimensional  numbers, 
i.e. the steepness and the relative spectral band width, $\Delta f/f_0$, are maintained.
Indeed, the JONSWAP spectrum is widely used for engineering applications: a typical experiment
consists in placing a structure (a model of a ship or an offshore platform) at some target 
 location in the tank, launching waves characterized by the JONSWAP spectrum 
 at one end and measuring the response of the structure \cite{goda00}. Such response is then analyzed for various 
 parameters in the JONSWAP spectrum.
 
\section{Experimental setups } \label{sec:setup}

\subsection{The hydrodynamical experiment}

\begin{figure}[h]
\includegraphics[width=8cm]{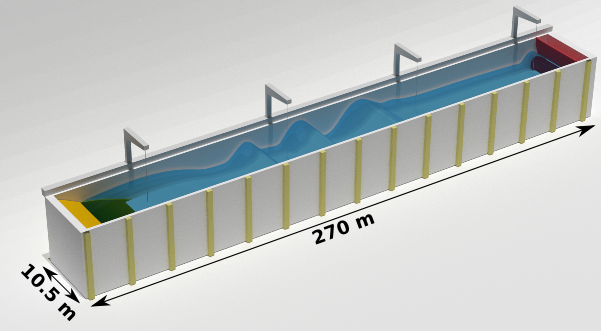}
\caption{Sketch of the  water wave tank used in the experiment. At one end a fully programmable 
wave maker is placed in order to generate waves; at the other end an absorbing beach is 
placed in order to minimize reflections. Fifteen probes were placed along the center of  the tank 
at distances of $10, 30, 35, 40, 45, 60, 65, 70, 75, 80, 85, 115,120, 160, 200$ meters from the wave maker;
two extra probes were placed transversally at $z=75$ meters and $z=160$ meters in order to verify if any transverse mode 
developed in the tank.}
\label{fig:setupw}
\end{figure}

The following description is taken from reference \cite{ONO06} where the experimental setup is described in detail.
The experiment has been performed at Marintek in Trondheim (Norway) in one of the longest existing wave tank. The length of the flume is 270 m and its width is 10.5 m (see figure \ref{fig:setupw}). The depth of the tank is 10 meters for the first 85 meters, then 5 meters for the rest of the flume. For waves of 1.5 seconds used in the present experiment, the water depth parameter $k_0 h\simeq 9$, with $k_0$ the wave number and $h$ the water depth, corresponds to the deep water regime.  A  wave-maker (flap type) located at one end of the tank was used to generate the waves.  A sloping beach is located at the far end of the tank opposite the wave maker so that wave reflection is minimized. The wave surface elevation was measured simultaneously by 19 probes placed at different locations along the flume; conductance wave gauges  were used. 

Once the parameters of the JONSWAP spectrum and the random phases, $\phi_i$  are selected, 5 different time series with a duration of 32 minutes each, have been produced as follows:
\begin{equation}
\eta(t)=\sum_{i=1}^N \sqrt{2 S(f_i)} \cos(2\pi f_i t+\phi_i).
\end{equation}
The time series are the input to the software that controls the wave-maker. 
Note that the dynamics is not statistically homogeneous in space but it is statistically stationary in time at each distance from the wave maker.

All experiments were performed with a nominal peak period of the JONSWAP spectrum 
of 1.5 seconds.

Different values of $\alpha$ and $\gamma$ were selected in order to have two different values of 
spectral bandwidth and significant wave height. In the Table \ref{table:parameters_water} 
we report the parameters of the 
 two experiments, run $A$ and run $B$, here considered.
\begin{table}[h!]
\begin{tabular}{|c|c|c|c|c|c|}
  \hline 
 &   $\alpha$  & $\gamma $ & $f_0$ (Hz) & $\Delta f/f_0$ &  $H_s$ (cm) \\
 \hline 
run $A$ & 0.0112 & 4.8  & 0.667&    0.178 & 14 \\
run $B$ & 0.0113 & 8 & 0.667 & 0.154 &  17   \\
\hline
\end{tabular}
\caption{Parameters of the JONSWAP spectrum for the two hydrodynamical experiments.}
\label{table:parameters_water}
\end{table}
 Here we specify that the parameters contained in the table are the one obtained 
 by a fitting of the spectrum measured at the first wave gauge, i.e. at 10 meters from the 
 wave maker (the parameters are slightly different from the {\it nominal} one that are
 specified to the programmable wave-maker).
 Therefore, experiments in optics and numerical computations will all start from a JONSWAP spectrum with random  phases with the parameters measured at  10 meters from the wave maker. With such choice we define the new {\it zero-coordinate}, $z=0$ of our experiment to be located at 10 meters from the wave maker.
  
  As mentioned in the Section \ref{sec:unifying}, for a correct comparison between the hydrodynamical and the optical experiments it is of paramount importance to introduce the ratio between  the linear to nonlinear lengths. Using Eqs. (\ref{eq:zlznl}), $\beta_{2,hydro}=2/g$, $\chi^{(3)}_{hydro}=k_0^3$ and the dispersion relation of deep water waves $\omega_0^2=g k_0$, one finds for the hydrodynamical experiment that linear and nonlinear lengths are :
  
\begin{equation}
z_{lin}=\frac{g}{\Delta \omega^2}=\frac{\omega_0^2}{k_0 \Delta \omega^2}=\frac{f_0^2}{k_0 \Delta f^2}
\label{eq:zllw}
\end{equation}

and

\begin{equation}
z_{nlin}=\frac{1}{k_0^3 \langle|A_0|^2\rangle}=\frac{8}{k_0^3 H_s^2},
\label{eq:znlw}
\end{equation}

therefore
\begin{equation}
\epsilon=\frac{z_{lin}}{z_{nlin}}=\frac{k_0^2 \langle|A_0|^2\rangle}{(\Delta f/ f_0)^2},
\label{eq:bfil}
\end{equation}
i.e. the ratio between the square of the steepness and the square of the spectral bandwidth of the initial condition, see \cite{ONO01}. Comparison between the optical and hydrodynamical experiments will be made by introducing the 
nondimensional coordinate $z=z'/z_{nlin}$ with $z'$ the distance expressed in meters from the first probe.
In Table \ref{table:parameters_water2}  the linear, the nonlinear lengths, their ratio and $z_{max}=z'_{max}/z_{nlin}$, with $z'_{max}$
the distance of the last probe from the first one, are reported. \begin{table}[h!]
\begin{tabular}{|c|c|c|c|c|}
  \hline 
&  $z_{lin}$ (m)  & $z_{nlin}$ (m) &$\epsilon=z_{lin}/z_{nlin}$ & $z'_{max}/z_{nlin}$ \\
 \hline 
run $A$  & 17.5 & 71.1 & $\sim$ 0.25 &2.67 \\
run $B$  & 23.5 & 48.2 &$\sim$ 0.50 & 3.94\\
\hline
\end{tabular}
\caption{ Length scales and nonlinear parameters for the two hydrodynamical experiments.}
\label{table:parameters_water2}
\end{table}

Considering that the dominant wave generated in the wave tank has a period of 1.5 seconds, then our statistics can be obtained at each location by averaging over 6400 linear time scales.
\subsection{The optical experiment}

 \begin{figure}[h]
\includegraphics[width=8cm]{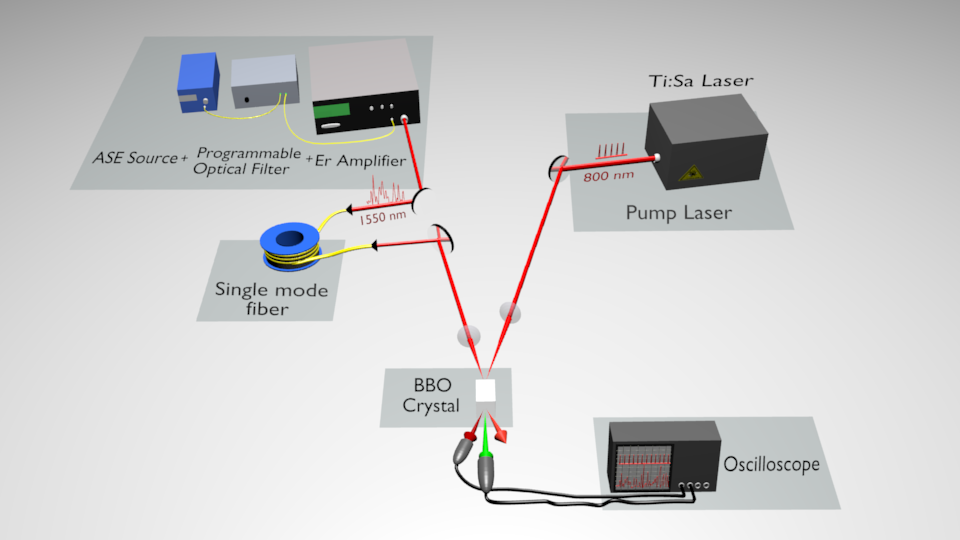}
\caption{Optical sampling of the partially-coherent wave
  fluctuating with time (the signal) is achieved from Sum Frequency Generation (SFG).
  Green pulses are generated at $\lambda=529$nm from
  the interaction of the signal with femtosecond pump pulses inside a $\chi^{(2)}$ crystal.
The $140$fs pump pulses are  emitted by  mode-locked laser at $\lambda_p=800$nm. The partially coherent  wave is emitted by an ASE source at $\lambda_s=1562$nm and is amplified by an Erbium fiber amplifier. Statistics of partially-coherent light is measured from the SFG process
  either directly at the output of the laser or after propagation
  inside an optical fiber.}
\label{fig:setup}
\end{figure}

The optical experimental setup is displayed in Fig. \ref{fig:setup}. It can be divided in three parts : i) the source of partially coherent waves, ii) nonlinear propagation in optical fiber and iii) optical sampling detection. \\

i) The random optical waves source is comparable to the one described in \cite{Suret:16}: the partially coherent light (i.e., the initial condition) is
generated by an Erbium fibre broadband Amplified Spontaneous Emission (ASE) source (Highwave), which is spectrally filtered (with adjustable shape and line width)
using a programmable optical filter (Waveshaper 1000S, Finisar). The central wavelength is $\lambda\sim1562$~nm corresponding to a carrier wave frequency $f_{0,optics}=192$THz.  The output light is then amplified by an Erbium-doped fiber amplifier (Keopsys).

Using the programmable optical filter, the optical spectrum of the partially coherent light emitted by the amplifier is precisely designed to assume a JONSWAP shape. Note that this step is not straightforward because of the amplification and several empirical feedback loops are necessary to converge. The red line in  Fig. \ref{fig:Spectra} is a typical spectrum measured with an Optical Spectrum Analyser at the output of the random source. Our setup allows us to obtain initial optical spectrum that is very close to the ideal JONSWAP spectrum (black dashed lines) and to the spectrum of water waves measured at the first gauge of the water tank (blue line). \\

ii) The amplified random light is launched into standard single-mode fibers (SMF28) having different lengths $z'_{fiber}$. The wavelength $\lambda\sim1562$~nm of random light falls into the  anomalous (focusing) regime of dispersion of the fiber that  has a group velocity dispersion coefficient $\beta_{2,fiber}=-22$~ps$^2$km$^{-1}$ and a  Kerr effect coefficient $\chi^{(3)}_{fiber}\simeq 1.3$~W$^{-1}$km$^{-1}$\\

iii) The statistics of optical power is measured at the input and at the output ends of the optical fiber with the optical sampling setup described in \cite{Walczak:15}. Sum frequency generation (SFG) between the random waves under study and  femtosecond pump pulses is obtained at a wavelength $\simeq 529$nm in a BBO crystal. The $140$fs-long pump pulses are emitted by a mode-locked Ti:Sa laser (Coherent Cameleon ultra II) at $800$nm with a repetition rate of $80$MHz. The BBO crystal has 8~mm length and is cut for noncollinear
type-I SFG  ($\theta = 24.2^\circ$, $\phi = 90^\circ$, external angle
between pump and signal  $= 12.5^\circ$). By ensuring that the typical
power of random waves ($\sim 2$W) is much weaker than the peak power
($\sim 4.10^5$W) of the  pump pulses, the energy of each SFG pulse is proportional to the instantaneous optical power $|A|^2$ carried by the random waves \cite{Boyd}. The short green pulses are observed by using a highly sensitive photodiode (MenloSystem FPD310-FV)  having  a gain of $\simeq 10^4$  and a rise
time of $0.7$ns. We record the output of  the photodiode  with a
fast oscilloscope (Lecroy WaveRunner 104MXi-A, bandwidth 1GHz,
10GS/s). We compute the PDFs  of $|A|^2$ from typical ensemble of
approximately 8 millions  measurements of SFG peak powers (see \cite{Walczak:15} for details).\\

The key point of the optical experiments  is to use similar values of the relevant parameters in optical fiber and water tank experiments. This means that for a given initial statistics of water waves and for a given length of the water tank, the reduced parameters $z=z'/z_{nlin}$ and $\epsilon=z_{lin}/z_{nlin}$ should be identical in the optical experiment. In order to compare the statistics of water waves measured at a given length of propagation in the water tank together with the statistics of optical waves at the output of a fiber of length $z'_{fiber}$, we have used the following methodology :

 i) $z=z'/z_{nlin}$ is computed by using Eqs. (\ref{eq:zlznl}) and (\ref{eq:P0ww}) and the parameters of the water tank experiments;  

 ii) the mean optical power $P_0$ is computed by using Eq. (\ref{eq:zlznl}) and the value of optical nonlinearity  $\chi^{(3)}_{fiber}$;

  iii) the full width at half maximum $\Delta f_{hydro}$ of the input JONSWAP water waves spectrum is computed from the Eq. (\ref{eq:jonswap}) by using the parameters of the water tank experiments;

 iv) by imposing the same value for $\epsilon$ [Eq. (\ref{eq:epsilon})] in optical and hydrodynamical experiments,  one obtains the value of the full width at half maximum $\Delta f_{optics}$ of the input JONSWAP  spectrum to be used in the optical experiment : 
 \begin{equation}
   \Delta f_{optics} =\mu  \Delta f_{hydro} \text{   with   } \mu =  4\; \sqrt{\frac{\chi^{(3)}_{fiber}\, P_0}{\beta_{2,fiber}\,g k_0^3H_s^2}}
   \label{eq:freqUnit}
 \end{equation}
Note that $\mu$ represents the ratio between the typical time scales of the structures emerging in water waves and optical fibers experiments and it is of the order $10^{12}$ in our experiments.

v) Finally, the spectrum $S_{optics}(f)$ of the partially coherent optical waves is designed  by using the programmable optical filter. The optical spectrum follows the JONSWAP spectrum given by the Eq. (\ref{eq:jonswap}) with the broadening factor $\mu$:
 \begin{equation}
   S_{optics}\big (f_{optics} \big )=S\bigg ( \frac{f_{optics}-f_{0,optics}}{\mu}+f_0 \bigg )
   \label{eq:jonswapOptics}
 \end{equation}
 where $f_0$ is the central frequency of water waves.\\

By using the procedure outlined in (i-v), for a given propagation
length in water tank and a given value of the significant water wave
height, an equivalent optical fiber experiment is performed. Contrary
to the so-called cutback
technique~\cite{Barviau:06,kibler2012observation, mussot2014fermi,
  Tikan:17}, the technique used here (and similar for example to the one used in
\cite{kibler2010peregrine}) allows  to explore various normalized
propagation lengths with only one optical fiber having a fixed length.  The key point is to adjust the spectral width and the optical power in order to change proportionally the number of linear lengths and of nonlinear lengths experienced by the waves.

 The set of parameters used in the optical fiber experiments and their counterparts in the water tank are displayed in tables \ref{table:param1} and \ref{table:param2} for run $A$ and run $B$ experiments, respectively. In principle, our scaling technique allows the use of one fiber length in order to investigate all the normalized lengths of propagation. However, in order to use easily available optical powers and to keep the signal to noise ratio roughly constant ({\it i.e.} by avoiding very low power), we have used several lengths of optical fibers.  The tables contain the lengths of the different fibers used in the experiments and the corresponding propagation distances in meters achieved in the water wave experiment (last column $z'_{hydro}$). For the same length of the fiber, for example 500 meters (see the fifth and sixth rows in Table  \ref{table:param1}), it is possible to explore a propagation distances equivalent to 20 meters in the equivalent water wave experiment by changing the width of the spectrum and its power that differs.



 \begin{table}[h!]
 \begin{tabular}{|c|c|c|c|c|c|}
  \hline
   $z'_{fiber}$~(m)    & $P_0$~(W) & $\Delta  f_{opt}$~(THz) & $z'/z_{lin}$& $z'/z_{nlin}$ &  $z'_{hydro}$~(m) \\
  \hline
  50   & 2.0 & 0.150 & 0.49 & 0.13& 10\\
  125   & 1.6 & 0.141 & 1.08& 0.26 & 20\\
  250   & 1.2 & 0.128 & 1.78 & 0.40& 30\\
  250   & 1.6 & 0.140 & 2.13 & 0.53 & 40\\
  500   & 1.2 & 0.129 & 3.61 & 0.79 & 60\\
  500   & 1.6 & 0.140 & 4.25 & 1.06 & 80\\
  1000  & 1.2 & 0.130 & 7.33 & 1.59 & 120 \\
  1250   & 1.2 & 0.117 & 7.42& 1.98 & 150\\
  1000   & 2.0 & 0.139 & 8.38& 2.64 & 200\\

  \hline
 \end{tabular}
 \caption{Optical parameters corresponding to the water wave experiment run $A$, see Tables \ref{table:parameters_water} and 
 \ref{table:parameters_water2}.}
\label{table:param1}
 \end{table}

 \begin{table}[h!]
 \begin{tabular}{|c|c|c|c|c|c|}
  \hline
   $z'_{fiber}$~(m)  & $P_0$~(W) & $\Delta f_{opt}$~(THz) & $z'/z_{lin}$ & $z'/z_{nlin}$ &  $z'_{hydro}$~(m) \\
  \hline
  125  &1.2 & 0.089 & 0.43& 0.20& 10  \\
  125  & 2.4 & 0.122 & 0.81 & 0.39 & 20  \\
  250  & 1.8 & 0.112 & 1.37 & 0.59 & 30  \\
  250 & 2.4 & 0.123 & 1.64 & 0.78 & 40   \\
  500  & 1.8 & 0.114 & 2.82 & 1.17 & 60\\
  500 & 2.1 & 0.108 & 2.53 & 1.37 & 70  \\
  500  & 2.4 & 0.127 & 3.50  & 1.56 & 80 \\
  1000  & 1.3 & 0.084 & 3.06 & 1.66 & 85 \\
  1000 & 1.8 & 0.114 & 5.62 & 2.34  & 120  \\
  1250 & 1.8 & 0.105 & 5.98 & 2.93  & 150 \\
  1000 & 3.0 & 0.138 & 8.26 & 3.90  & 200 \\

  \hline
 \end{tabular}
  \caption{Optical parameters corresponding to the water wave experiment run $B$, see Tables \ref{table:parameters_water} and 
 \ref{table:parameters_water2}.}
\label{table:param2}

\end{table}

 \section{Results and Discussion} \label{sec:results}

\subsection{Input spectra}
\begin{figure}[h]
\includegraphics[width=8cm]{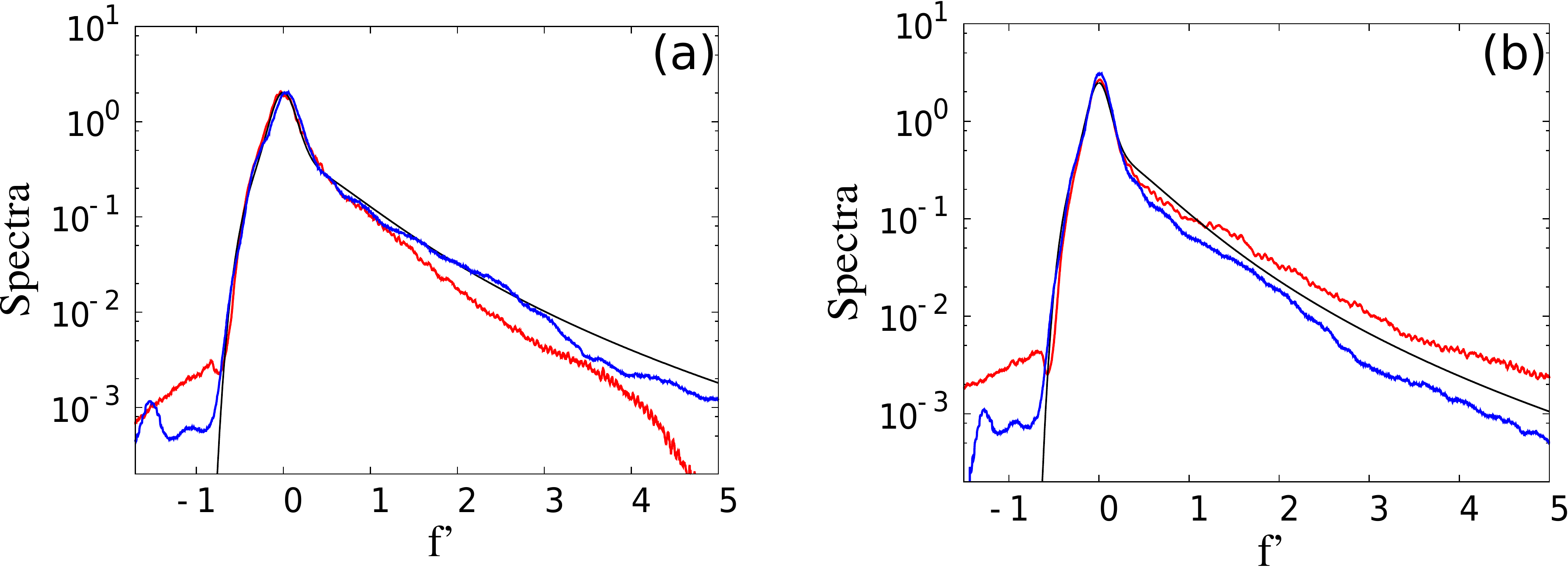}
\caption{ Spectral power density at $z=0$ in the  optical fiber (red
  line) and in the water tank (blue line) for run $A$ (a) and $B$ (b). The JONSWAP spectrum (black dashed line) with parameters taken from Table \ref{table:parameters_water} is also shown. The frequency $f'$ is normalized by using Eq. (\ref{eq:norm_freq})}
\label{fig:Spectra}
\end{figure}

An important step towards a comparison between the results is to start the hydrodynamical and 
optical experiment with compatible spectra.  In order to check such compatibility, we find useful to introduce the following nondimensional frequency as follows:
\begin{equation}
f'=\tau (f-f_0),
\label{eq:norm_freq}
\end{equation}
where $\tau$ is defined in equation (\ref{time}). In optics  $\tau_{optics}=\sqrt{|\beta_2|/ (2 \chi^{(3)} P_0)}$, 
while in hydrodynamics  $\tau_{hydro}=\sqrt{8/(g k_0^3 H_s^2)}$.
In figure (\ref{fig:Spectra}a) and (\ref{fig:Spectra}b) we show the frequency spectra as a function of $f'$ for run $A$ and run $B$ at 
$z'=0$; the JONSWAP spectrum with parameters reported in Table \ref{table:parameters_water} is also shown. 
A very good agreement in the initial condition is shown for the most energetic part of the spectrum. 

\subsection{Statistical properties of the intensity}
\subsubsection{The probability density function of the Intesity}
As mentioned in Section \ref{sec:statistics} we concentrate our analysis on the statistical properties of  the 
normalized intensity of the wave field: $I=|A|^2/P_0$ in optics and $I=|A|^2/(2\sigma^2)$  in hydrodynamics. We recall that the surface elevation is measured in the hydrodynamical experiment. Therefore, in order to compute the intensity, the envelope $A$ has to be calculated using the  Hilbert transform.

\begin{figure}[h]
\includegraphics[width=9cm]{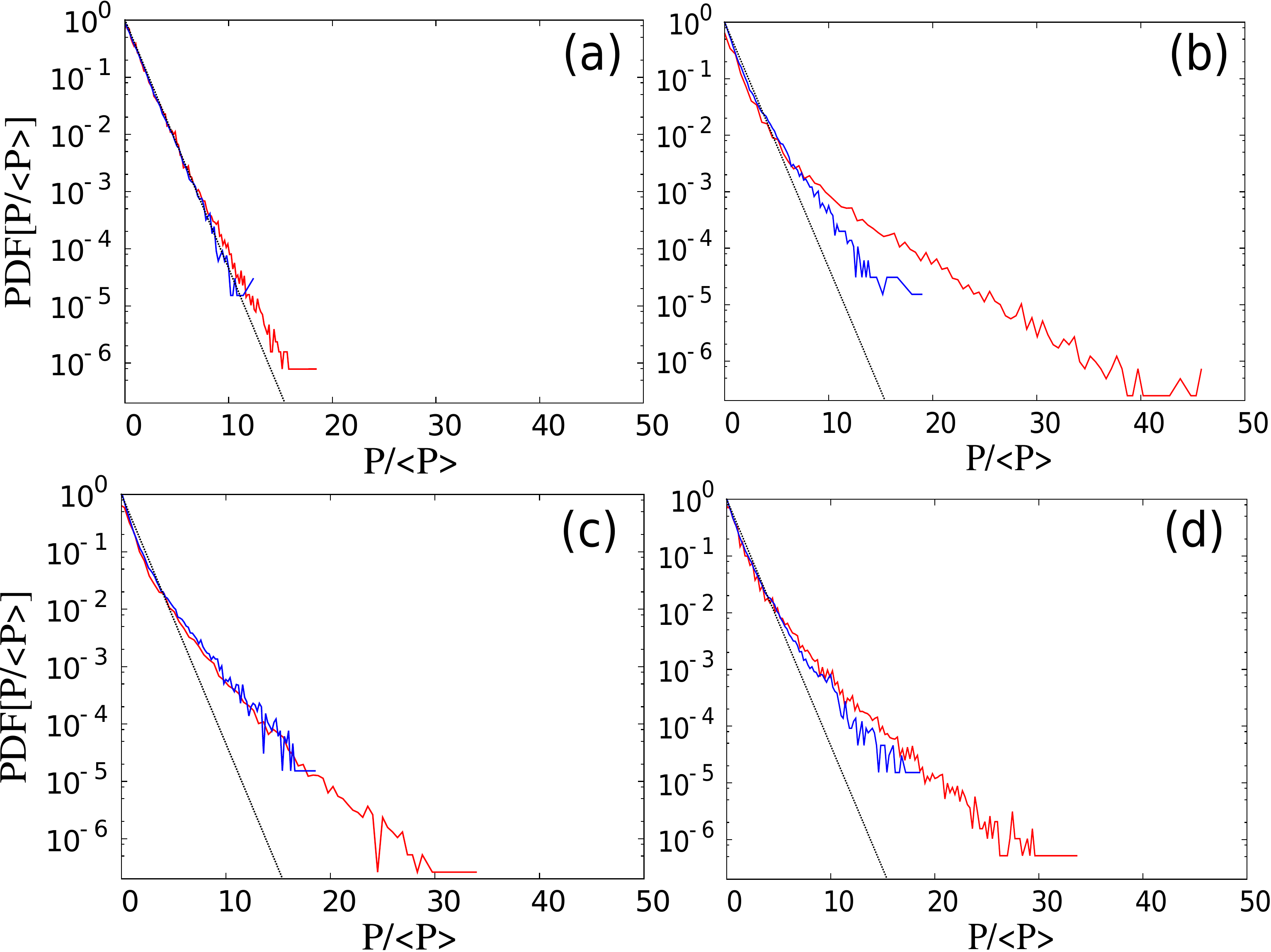}
\caption{Probability Density Function  of the normalized Intensity
    $I$ for the experiments in run $B$ ($\epsilon \sim 0.5$ at different nondimensional distances: a) $z=z'/z_{nlin}=0$ b) $z\simeq0.6$ c) $z\simeq 1.4$ d) $z\simeq 4.$.
 The blue line corresponds to the water wave experiment and the red one to the optical one. The black dashed line corresponds to the  Exponential distribution.}
\label{fig:PDF}
\end{figure}

The Probability Density Functions (PDF) of $I$ measured both in the water wave tank and in the optical fiber experiments for a strength of the nonlinearity corresponding to  run $B$ are displayed in Fig. \ref{fig:PDF} for different propagation lengths.  As expected, the PDFs of the intensity measured at the beginning of the fiber/tank ($z_{fiber}=z_{hydro}=0$) are very close to the Exponential distribution. This is expected because the waves are generated by using a prescribed spectral shape with random phases. When the length of propagation is sufficiently large, the nonlinearity starts to become important. The statistics of both optical and hydrodynamical intensities deviate from the Exponential distribution, displaying heavy tails. Strikingly, for high values of the normalized length (typically $z>2$), the PDFs measured in optical fibers and in the water wave tank experiments are remarkably close (see Fig.  \ref{fig:PDF}c and  \ref{fig:PDF}d). For intermediate values of $z\simeq0.5-1.$, the optical waves exhibit stronger deviation from the Exponential distribution than the water waves (see Fig.  \ref{fig:PDF}b). This issue will be further discussed in the Section C where numerical simulations are reported. 
It is interesting to note the difference probability levels that are achieved in optical fibers and the water tanks. For the hydrodynamics experiments, PDFs are obtained by averaging over five realizations with different phases each lasting 32 minutes (in between the different realizations, about 30 minutes were needed in order to let the waves damp in the tank). The time needed in the optical experiment to collect the data for a PDF is less than 10~ms and the probability level achieved is more than one order of magnitude lower that in the water wave experiment.

The PDFs of run $A$ (not reported here) display the same features of those of  run $B$, with the only difference being that the level of nonlinearity is lower and the deviations from the Exponential distribution are less prominent. 
 
\subsubsection{The normalized fourth-order moment of the probability density function of the envelope}
\begin{figure}[h]
\includegraphics[width=8.7cm]{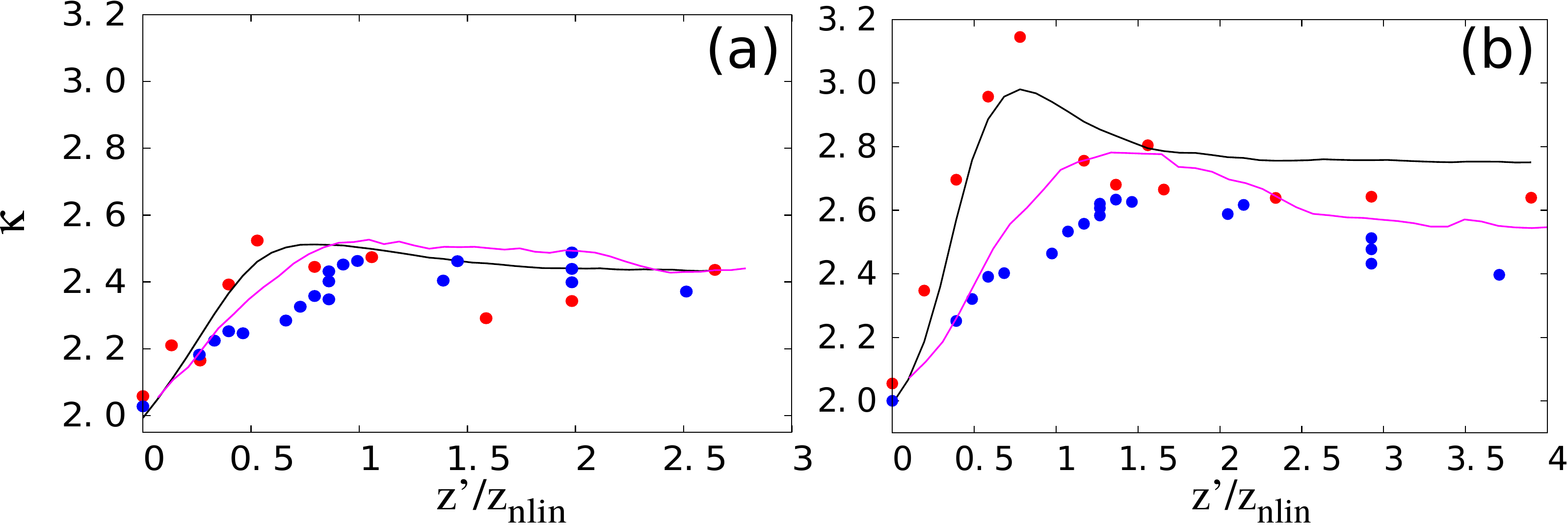}
\caption{
Fourth-order normalized moment of the wave envelope moment, 
$\kappa=\langle |A|^4 \rangle / \langle |A|^2 \rangle ^2$, as a function of the nondimensional propagation coordinate $z$: experiments and numerical simulations for run $A$ (panel a) and run  $B$ (panel b). Optical experiments (red points), hydrodynamical experiments (blue points), simulations of the NLSE (black lines), simulations of Euler equation (magenta line).
}
\label{fig:kurtosis}
\end{figure}

In order to compare the statistical properties of water and optical waves, we have computed the normalized fourth order moment $\kappa$, see Eq. (\ref{eq:kappa}), of the wave envelope for each experimental parameters as a function of the normalized propagation distance $z=z'/z_{nlin}$.
We remark that values larger than 2 imply a departure from linear predictions. 
In  Fig. \ref{fig:kurtosis}, the fourth-order moment of $|A|$ is plotted as a function of $z$ for hydrodynamics (blue dots) and optical (red dots) tests: the case corresponding to run $A$ is displayed on the Fig.  \ref{fig:kurtosis}a and the higher nonlinear case,  run $B$, is displayed on the Fig. \ref{fig:kurtosis}b.\\

Numerical simulations of the NLS equation and the Euler equation for  water waves are also displayed (see Section \ref{sec:numer_sim}).

For both runs, $A$ and $B$, $\kappa$ starts from the value of 2 and then grows on a spatial scale of the order of 
a nonlinear length. Run $B$ is more nonlinear than run $A$ and displays larger deviations from the linear predictions.
The optical experiment displays a faster growth of $\kappa$ and it is more consistent with the NLE equation. Both experiments show 
a maximum value of $\kappa$ at some length after which $\kappa$ is almost constant.
A significant overshoot in the evolution of $\kappa$ for optical waves around $z=0.7$ in run $B$  (Fig. \ref{fig:kurtosis}b) can be observed, while it is less pronounced for water waves.

In the weaker nonlinear case, run $A$, displayed in Fig. \ref{fig:kurtosis}a, the agreement between optical and water waves experiments is strikingly good. The evolution of $\kappa$ is a little bit slower for water waves but both cases seem to reach a stationary state $\kappa\simeq2.4$ around $z=1$. Note that the point $z=1.5$ has been recorded with a SMF fiber of a different spool and we interpret the decreasing of $\kappa$ for this point as the influence of a slight change in the Kerr effect coefficient.

Finally, we want to emphasise that, despite there are some differences between the optical and 
 hydrodynamical results, the agreement between the two completely different experiments is remarkable. 
 
\subsection{Numerical Simulations} \label{sec:numer_sim}
In order to interpret the differences between optical and waters waves statistics, we have performed numerical simulations using two different models: the NLSE and the Euler equations for water waves has been solved numerically. 

\subsubsection{NLSE simulations}
In standard optical fibers, the power losses are very low  ($\simeq 0.18$dB/km {\it i.e.} $2\%$ in $500$m for example) and the experiments performed with partially coherent waves having a narrow spectrum and  a central wavenlength far from the zero dispersion wavelength are known to be well described by the  NLSE~\cite{Agrawal,Suret:16}. Note that mean powers used in optical experiments are relatively low and that we have checked from optical spectra measurement that stimulated Raman scattering can be neglected. 

The equation (\ref{nls}) has been solved numerically using periodic boundary condition in time. We have considered a  complex field $A(z=0,t)$ having a JONSWAP optical power spectrum, Eq. (\ref{eq:jonswapOptics}), 
and random phases. Statistical properties of the waves have been computed from Monte Carlo simulations 
made with an ensemble of $10^4$ realizations characterized by different random phases for the initial condition. The normalized fourth-order moment $\kappa$ is plotted with black lines in the Fig. \ref{fig:kurtosis} together with the optical fiber and water tank experiments.

The agreement between experiments performed in optical fibers and numerical simulations is overall very good.  

Concerning the comparison of the simulations with the hydrodynamical  experiment, it is clear especially from Fig. \ref{fig:kurtosis}b that  the dynamics of water waves is slower than the numerical predictions. This result is well known and already documented for example in \cite{Zhang2014a} and it is basically due to the fact that the spectrum is not as narrow as implied by the NLS theory. Moreover, white capping has also been observed in the 
experiment; such phenomenon is not reproducible within the NLS theory.

In conclusion, the evolution of $\kappa$ as a function of $z=z'/z_{nlin}$ computed from NLSE reproduces better the optical rather than the hydrodynamical experiment. This is not a surprise because the optical fiber has been designed in order to be properly described by the NLSE, while there is no possibility to modify the medium in order to change the dispersion and the nonlinearity to match better the equation in the water wave context.

\subsubsection{Euler  simulations}
In Fig. \ref{fig:kurtosis} we have also reported numerical simulations of the Euler equations for water waves. The fluid has been considered as inviscid, incompressible and irrotational. The set of equations for the surface elevation $\eta(x,t)$ and for the velocity potential, $\psi(x,t)$, calculated on the free surface that have been solved are the following:
\begin{equation}
\begin{split}
&\frac{\partial \eta} {\partial t}+
\frac{\partial \psi} {\partial x}\frac{\partial \eta} {\partial x}
- w\bigg[1+\left(\frac{\partial \eta}{\partial x}\right)^2 \bigg]=0\\
&\frac{\partial \psi} {\partial t}+g \eta+
\frac{1}{2}\bigg(\frac{\partial \psi} {\partial x}\bigg)^2
-\frac{1}{2} w^2\bigg[1+\left(\frac{\partial \eta}{\partial x}\right)^2 \bigg]=0,
\end{split}
\end{equation}
where $w$ is the vertical velocity computed on the free surface. Its calculation in principle
requires the knowledge of the velocity field  in the whole domain (under the free surface). In order to 
address this problem we have used the so called Higher Order Spectral Method (see \cite{dommermuth1987high,west1987new}) by which an iterative procedure is used for expressing the vertical velocity as a function of the surface elevation and the velocity potential on the surface. When the iterative
scheme is truncated adequately the system is Hamiltonian. 
The system is solved as  evolution equations in time with periodic boundary condition in space. In order to compare the results with the experiments, the group velocity is then used to convert time to space. The numerical method does not allow for wave breaking. The initial conditions are provided by the JONSWAP spectrum (once converted in a wave numbers spectrum) with random phases for the surface elevation and the velocity potential is then calculated assuming linear theory (see \cite{toffolijfm10} for details). 

The results obtained from the simulations are displayed in Fig. \ref{fig:kurtosis}. The evolution of $\kappa$ provided by the Euler equations shows a much better agreement to the hydrodynamical experiment with respect to the NLSE simulations. The values of $\kappa$ are just slightly overestimated with respect to the experimental results, especially for run $B$. We believe that such effect is largely due to the impossibility of the model to reproduce the wave breaking phenomenon visually observed in the tank. Any  damping such as the one due to the presence of lateral walls is also not included in the simulations. The asymptotic state reached in the simulations is slightly larger than the one obtained in the experiments.
 
\section{Conclusions} \label{sec:conclusions}
Starting from the pioneering work of Solli et al. \cite{solli2007optical}, the concept of rogue waves has been introduced in the optical community. Since then, a lot of work has been done, trying to deepen the analogies between optical and hydrodynamical rogue waves  \cite{Akhmediev:16}. The bridge for the analogy finds its roots in the universality of the NLS equation \cite{chabchoub2015nonlinear} and its capability of describing weakly nonlinear dispersive waves in different contexts. Indeed, exact breather solutions of the NLSE have been reproduced with some degree of success both in hydrodynamics and optics \cite{kibler2015superregular}. However, apart from very special conditions, ocean waves cannot be considered as a small perturbation of a coherent wave. Measurements during field experiments of the surface elevation show that ocean waves are characterized by a finite-width spectrum whose phases are hardly distinguishable from a set of random phases.

In the present paper an optical experiment has been  devised in order to properly reproduce the statistical properties of gravity waves measured  in a long wave tank with initial conditions characterized by a JONSWAP spectrum and random phases. The key role for designing properly the optical experiment is played by the NLSE. Indeed, out of the equation nondimensional distances and nonlinear parameters can be derived and used for comparing the experiments. Two sets of experiments have been performed characterized by two different ratios between linear and nonlinear propagation distances. In the context of ocean waves, such ratio is nothing but the square of the so called Benjamin-Feir Index defined in \cite{ONO01,janssenbook04}. Larger values of such index in the initial condition leads to the formation of more rogue waves than the linear theory would predict and consequently the formation of heavy tails in the probability distribution of the wave intensity. Such behaviour has been observed in the optical experiment.

 A one to one comparison between the hydrodynamical and the optical experiments has been performed. The focus has been on the probability density function of the wave intensity and on the evolution of the fourth-order moment of the probability density function of the wave envelope. While in a wave tank measurements of the surface elevation along the tank are always possible, they can be done only at its end in an optical fiber. An appropriate technique based once more on the NLSE equation has allowed us to perform measurements at different propagation distances using the same fiber by changing the width of the initial spectrum and its power. 

Having in mind that the optical and hydrodynamical experiments are completely different,  a quantitative comparison between them shows a remarkable agreement. Heavy tails and deviation from linear statistics are observed in both experiments at common nondimensional propagation distances. Faster evolution of the dynamics is observed in the optical experiment. As mentioned, this is due to the fact that the latter has been designed to match properly the NLSE equation which is not an optimal model for water waves.  Numerical simulations of the NLSE confirm our findings. A numerical study of the Euler equation has been performed and the results show a better agreement than the NLSE with the hydrodynamical experiment.  Here we stress that propagation in both experiments is one dimensional. Recently, it has been found that two-dimensionality may play an important role in the statistical properties of ocean waves \cite{ONO09,waseda09,Nobuhito11}. We hope that our success in the comparison will trigger new work in the even more complicated optical set-ups ruled by  two-dimensional propagation. \\

{\bf Acknowledgments}

This work has been partially supported by the French Agence Nationale de la Recherche through the LABEX CEMPI project (ANR-11-LABX-0007) and the OPTIROC project (ANR-12-BS04-0011 OPTIROC) as well as by the Ministry of Higher Education and Research, Nord-Pas de Calais Regional Council and European Regional Development Fund (ERDF) through the Contrat de Projets Etat-Re´gion (CPER Photonics for Society P4S).
M.O. has been funded by  Progetto di Ricerca d'Ateneo CSTO160004 . The
authors are grateful to E. Courtade, F. Anquez, J. Pesez  and the
Biophysics of Cellular Stress Response group of the PhLAM for
providing the femtosecond laser used in optical sampling. The authors
acknowledge G. Genty for fruitful discussions. Dr. B. Giulinico is acknowledged for discussion.

\end{document}